\documentclass[11pt]{article}
\usepackage{amsmath}
\usepackage{amsfonts}
\usepackage{amssymb}
\usepackage{graphicx}

\def\be\begin{equation}
 \def\ee{\end{equation}}
\def\bea{\begin{eqnarray}}
\def\eea{\end{eqnarray}}

\begin{document}
\begin{center}
\LARGE {\bf On the Anisotropic Big Bang Cosmology With A General
Non-Linear EoS  }
\end{center}
\begin{center}
{\bf  Kh. Saaidi}\footnote{ksaaidi@uok.ac.ir},
{\bf  S. Ziaee}\footnote{S.ziaee@uok.ac.ir}\\

{\it Department of Physics, Faculty of Science, University of
Kurdistan,  Sanandaj, Iran}

\end{center}
 \vskip 1cm
\begin{center}
{\bf{Abstract}}
\end{center}
We study the anisotropic Big Bang cosmology for a general non-linear
equation of state (EoS) in standard general relativity cosmology in early times.
 According to brane world cosmology and loop quantum gravity which introduce
 a quadratic term in the energy density, we
study a non-linear EoS with a general nonlinear term in the
energy density. We show that this general non-linear term
isotropize the universe at early times. And also, we find out
this model has a phantom like behavior for special case.


\newpage

\section{Introductions}

Standard Big Bang cosmology is based on three principal. The first
is the supposition that universe is homogeneous and isotropic on
large scales. The second is the supposition that the dynamics of
space-time is expanding by the Einstein field equations. The third
basis of the theory is that matter can be explained as a super
position of two classical perfect fluid, a radiation fluid with
relativistic EoS and pressure-less (cold) matter {\cite{1}}. In
recent decades, the observations of high redshift type $Ia$
supernova {\cite{2}} have shown that our universe is undergoing on
accelerated expansion instead of decelerated expansion
{\cite{3,4,5}}. Meanwhile, observation of cosmic microwave back
ground radiation (CMBR) {\cite{6,7,8}}, and large scale structure
{\cite{9}} show that the space is highly homogeneous and isotropic
on large scale. All this result strongly propose that, the
universe is permeated smoothly by dark energy. The dark energy and
accelerating universe have been discussed extensively from
different point of view {\cite{10,11,12}}. As already mentioned,
the large scale structure of the universe seems to be very
homogeneous and isotropic. However, observing on small scales, the
isotropy and homogeneity break down. In fact, the initial state of
the universe possesses some inhomogeneities, anisotropies and a
rather chaotic distribution of particles. Hence, it is tempting to
investigate more general inhomogeneous and anisotropic models in
early times, which should describe, as a consequence of their
evolution, the currently observed large scale structure together
with the isotropy limits observed in the CMBR, in x-ray
backgrounds (e.g. quasars at high redshift), and in number counts
in faint radio sources {\cite{13}}. In general relativity, without
the help of a cosmological constant or inflation, Collins and
Hawking {\cite{14}} tested the question (homogeneity and isotropy
of universe) in terms of an initial conditions analysis. They
achieved that the set of spatially homogeneous cosmological models
approaching isotropy in the limit of infinite times is of measure
zero in the space of all spatially homogeneous models. This in
turn implies that the isotropy of the models are unstable to
homogeneous and anisotropic disturbance. However, their
explanation of isotropization demands asymptotic stability of the
isotropic solution. An asymptotic stability analysis of Bianchi
models in general relativity {\cite{15}} show, for example that
in the Bianchi type I the anisotropy will not exactly vanished but can be bounded.\\
In the brane world scenario {\cite{16,17,18,19}} the extra
dimensions product a quadratic term in the energy density in the
effective 4-dimensional energy-momentum tensor. Under the logical
suppositions, this non-linear term has the very appealing effect
of suppressing anisotropy in early enough times. Hence, there is
a probability to suppressing anisotropy at the initial
singularity problem by adding a non-linear term to EoS
{\cite{20}}. For linear EoS $P=\omega\rho$ with $\omega=constant$,
in the case $\omega<-1$ the fluid behavior is a phantom manner
{\cite{21}}. In this case the energy density grows in the future
and decays in the past, so shear components is dominated in the
initial singularity. On the other hand, in coming closer
singularity matter with linear EoS $\omega<1$ is un-influential,
hence in this case singularity could be termed $velocity$
$dominated$ {\cite{22}}, but when $\omega>1$, the so called
super-stiff fluid (Ekpyrotic fluid), the energy density is
dominated in the early
enough times, so the initial singularity is isotropic.\\
In this paper we investigate the effects of a general non-linear
of EoS, which is a binary mixture of perfect fluid matter and
dark energy, in the Big Bang cosmological model. We want to
consider on the possibility use of a general non-linear term of
EoS as an effective manner of representing a dark energy, to
suppressing anisotropy when the singularity is approached. Here,
we indicate one of the most interesting results from adding
general non-linear term to EoS, is that the initial singularity,
become isotropic in contrast with ordinary cosmology. Also, we
show that for some special case this theory has phantom like
behavior.

\section{Non-linear EoS}

At first, we will review some models with non-linear term that
they have studied in this context. In the brane world scenario,
the our universe is self accelerating caused by an additional term
appearing in the Friedmann equation when constrained to the brane.
For a higher dimensional cosmology in the Randall-Sundrum
framework {\cite{23,24}} the Friedmann equation is obtained as
\begin{equation}
H^2={\frac{\Lambda_4}{3}}+{\frac{8\pi}{3{M_p}^2}}\rho+\epsilon{\left(\frac{4\pi}{3{M_5}^3}\right)^2}\rho^2+\frac{c}{a^4},
\end{equation}
where $\Lambda_4$ is the cosmological constant, $\epsilon=\pm1$, and $c$ is an integration constant whose magnitude as
good as sign depend on the initial conditions.\\
The geometric effects of loop quantum cosmology is that predict a $-\rho^2$ modification to the Friedmann equation {\cite{25,26}}. This modification is relevant in the very early universe. The effective Friedmann equation becomes
\begin{equation}
H^2={\frac{8\pi G\rho}{3}}\left(1-\frac{\rho}{\rho_c}\right)+\frac{\Lambda}{3},
\end{equation}
where $\rho_c=\frac{\sqrt{3}}{(16\pi^2\gamma^2)\rho_{pl}}$
{\cite{27}}. The $\rho^2$ modification is shared with the brane
world model, and some comparison works can be found in
{\cite{28,29,30}}. In this work, we introduce a class of
non-linear EoS and show that these models solve the anisotropy
problem in the early times universe. The general form of the EoS
which it include a non-linear term is introduced as
\begin{equation}
P=P_m+P_d=\omega\rho+\beta\frac{\rho^\alpha}{\rho_c^{(\alpha-1)}},
\end{equation}
where $\omega$ is a pure number, $\rho_c>0$ is a constant which
has energy dimension, $\beta$ is the constant parameter to show
the sign of the non-linear term. We can interpreting Eq.(3) as
a binary mixture of a perfect fluid and a dark energy with
positive pressure, for the case $\beta>0$.
\par Bianchi type I is the simplest approach for investigation of behavior of anisotropy
 at early times, which it is a subclass of the Bianchi class-$A$ {\cite{20}}.
The Bianchi type I cosmology can be determined by the Hubble
expansion scalar and the tracefree shear tensor $\sigma_{ij}$ in
which $i,j=1,..,3$ and $\sigma=({\frac{1}{2} \sigma_{ij}
\sigma^{ij}})^\frac{1}{2}$. The energy conservation equation for a
cosmological model including a perfect fluid is
\begin{equation}
\dot{\rho}=-3H(\rho+P),
\end{equation}
where $P$ is the pressure, $\rho$ is the energy density and $H$
is the Hubble parameter, i.e. $H=\frac{\dot{R}}{R}$. Using the
Bianchi type I model and Einstein equations (with suppose
${\frac{8{\pi}G}{c^4}=1}$), we can obtain

\begin{equation}
H^2=\frac{1}{3}(\rho+\sigma^2),
\end{equation}

\begin{equation}
\dot{H}=-\frac{1}{2}\left(\rho+P+2\sigma^2\right),
\end{equation}

\begin{equation}
\dot{\sigma}=-3H\sigma.\\
\end{equation}

\par With the help of non-linear EoS (3) and the energy conservation (4),
 one can obtain energy density as a function of scale factor.
\begin{equation}
\rho=\rho_c\left[
\frac{(1+\omega)B^{(\alpha-1)}}{R^{3(1+\omega)(\alpha-1)}-\beta
B^{(\alpha-1)}} \right]^{\frac{1}{\alpha - 1}},
\end{equation}
\begin{equation}
B^{(\alpha-1)}=\frac{\left({\rho_0}R_0^{3(1+\omega)}\right)^{\alpha-1}}{(1+\omega)\rho_c^{(\alpha-1)}+\beta{\rho_0^{(\alpha-1)}}},\\
\end{equation}
where $R_0$ and $\rho_0$ indicate the scale factor and energy
density at an arbitrary time $t_0$. This is acceptable for all
values of $\beta$, $\rho_0$ and $\omega$  except for $\omega=-1$.
Also we introduce
${\rho_\Lambda}^{(\alpha-1)}:=-\beta(1+\omega){\rho_c}^{(\alpha-1)}$,
that is an effective cosmological constant (for details see
reference {\cite{31}}). By using Eq.(7), we can obtain shear
scales as a function of the scale factor.
\begin{equation}
\sigma=\sigma_0 \left( \frac{R}{R_0} \right)^{-3}.\\
\end{equation}
According to above equation we realize that, the shear decreases
in an expanding model. Therefore, to eliminate the problem of
anisotropy at the early times, the initial singularity should be
dominated by the energy density. In fact the contribution of shear
scale is very smaller than energy density value. Also, we can
conventionally define
\begin{equation}
\Sigma=\frac{\sigma}{H},
\end{equation}
as a  dynamically scales (dimensionless) of anisotropy. Therefore,
if $\Sigma\rightarrow0$,  we can have a cosmological isotropic
model. For the disappearance the energy scale $\rho_c$ from the
dynamical system and simplify the analysis of equations, we
introduce the energy density scale, the shear scale and Hubble
parameter as dimensionless variable

\begin{equation}
\Omega=\left(\frac{\rho}{\rho_c}\right)^{\alpha-1},\quad
\kappa={\frac{\sigma}{\rho_c{^{\frac{1}{2}}}}},\quad
\eta=\frac{H}{\rho_c{^{\frac{1}{2}}}},
\end{equation}
where $\Omega$ is the normalized energy density, $\kappa$ is the
normalized shear scale, $\eta$ is the normalized Hubble
parameter.\\
Considering the above dimensionless variables, we can rewrite
Eqs.(10,11) as follows
\begin{equation}
\kappa=\kappa_0\left({\frac{R}{R_0}}\right)^{-3},
\end{equation}
\begin{equation}
\Sigma=\frac{\kappa}{\eta}.
\end{equation}
Also, here we define the effective cosmological constant point by
\begin{equation}
\Omega_\Lambda=\left({\frac{\rho_\Lambda}{\rho_c}}\right)^{\alpha-1}=-\beta(1+\omega).\\
\end{equation}
Note that in this paper we just investigate the region of the
state space in which the $\Omega\geq0$ and models which
$\eta\geq0$.

\section{Modified Polytropic Like Gas}

In this section, we want to study one example which EoS of it is
non-linear for considering the anisotropic behavior of the early
universe. Here, for simplicity, we rescale $\rho^{(\alpha-1)}$ as
$\frac{{\rho_c}^{(\alpha-1)}}{\beta}$. This is equivalent with the
case which $\beta=1$. For $\alpha=1+\frac{1}{n}$, Eq.(3) shows the
EoS of modified Polytropic like gas as
\begin{equation}
P=\omega\rho+\frac{\rho^{1+\frac{1}{n}}}{\rho_c^{\frac{1}{n}}},
\end{equation}
where $n$ index is a positive $(n>0)$. From Eq.(8) we arrive at
\begin{equation}
\rho=\rho_c\left[
\frac{(1+\omega)B^{\frac{1}{n}}}{R^{\frac{3}{n}(1+\omega)}-
B^{\frac{1}{n}}} \right]^{\frac{1}{\alpha - 1}}.\\
\end{equation}
By using Eq.(9) and
${\rho_\Lambda}^{\frac{1}{n}}:=-(1+\omega){\rho_c}^{\frac{1}{n}}$
we have
\begin{equation}
B^{\frac{1}{n}}=\frac{R_0^{{\frac{3}{n}}(1+\omega)}}{1-{\left(\frac{\rho_\Lambda}{\rho_0}\right)}^{\frac{1}{n}}}.\\
\end{equation}
Note that, in this paper the main difference is that the
singularity happen in the past if the following condition is
established
\begin{equation}
t\rightarrow{t_s},{\quad}R\rightarrow{R_s},{\quad}\rho\rightarrow\infty\quad
and{\quad}|P|\rightarrow\infty,
\end{equation}
where $t_s$ and $R_s$ are constant with $R_s\neq0$
{\cite{20,31,32}}. We can rewrite Eq.(17) for the energy density
in three different manner, introducing
$R_s=|B|^\frac{1}{3(1+\omega)}$ and
supposing $\rho>0$.\\

$(i):$ $(1+\omega)>0$,\quad ${\rho_\Lambda}^{\frac{1}{n}}<0$,\quad
${B^{\frac{1}{n}}}>0$.\\

In this case Eq.(17) will become
\begin{equation}
\rho=\rho_c\left[\frac{(1+\omega)}{(\frac{R}{R_s})^{{\frac{3}{n}}(1+\omega)}-1}\right]^n.\\
\end{equation}
According to Eq.(20) it is seen that for $\rho>0$, $R$ and $\rho$
must be in $(R_s,\infty)$ and $(0,\infty)$ respectively. By
making use of (20) and (19), one can realize that there is a
singularity at early times and the energy density, by increasing
scale factor $R$, decreases. This show that the linear term of EoS
is dominated. Hence, the fluid behavior is as a fluid with a linear EoS.\\

$(ii):$ $(1+\omega)<0$,\quad
$0<{\rho_\Lambda}^{\frac{1}{n}}<\rho^{\frac{1}{n}}$,\quad
${B^{\frac{1}{n}}}>0$.\\

For these parameter, Eq.(17) is written as
\begin{equation}
\rho=\rho_\Lambda\left[\frac{1}{1-(\frac{R_s}{R})^{{\frac{3}{n}}|1+\omega|}}\right]^n.\\
\end{equation}
In this case, one can see that $\rho$ must belongs to $(\infty,
\rho_\Lambda]$ also $R\in(R_s, \infty)$. So, the fluid close to a
singularity in the past (${R}\rightarrow{R_s}$,
$\rho\rightarrow\infty$), and at late times close to an effective
cosmological constant
($R\rightarrow\infty$, $\rho\rightarrow\rho_\Lambda$).\\

$(iii):$ $(1+\omega)<0$,\quad
$0<{\rho}^{\frac{1}{n}}<{\rho_\Lambda^{\frac{1}{n}}}$,\quad
${B^{\frac{1}{n}}}<0$.\\

In this case Eq.(17) is as
\begin{equation}
\rho=\rho_\Lambda\left[\frac{1}{1+(\frac{R_s}{R})^{{\frac{3}{n}}|1+\omega|}}\right]^n.\\
\end{equation}
It is seem that $R$ can varied between zero and infinity and then
$\rho$ is belonging to $(0,\rho_\Lambda)$. In this case fluid
behaves as phantom and it is close to an effective cosmological
constant at late times, i.e. ($R\rightarrow\infty$,
$\rho\rightarrow\rho_\Lambda$). Note that in cases $(i)$ and
$(ii)$, when the amount of energy density close to infinity, the
singularity happen at a finite scale factor, i.e.
${\rho}\rightarrow{\infty}$, ${R}\rightarrow{R_s}$.
\par Now we want consider the behavior of anisotropy for the EoS of
modified polytropic like gas. Using Eqs.(20-22) and (12-15), we
want to consider behavior of the shear scales and energy density
scales for three cases of fluid.\\

$(i):$ $(1+\omega)>0$,\quad ${\rho_\Lambda}^{\frac{1}{n}}<0$,\quad
${B^{\frac{1}{n}}}>0$,\\
\begin{equation}
\kappa=\kappa_0\left({\frac{\Omega}{(1+\omega)+\Omega}}\right)^{\frac{n}{1+\omega}},
\end{equation}
in this case, we have $\Omega\rightarrow\infty$ in early times,
so according to above equation the shear scale tends to a constant
value ($\kappa\rightarrow{\kappa_0}$). Therefore, the initial
singularity is isotropic, because in this case the initial
singularity is dominated by the energy density, so
$\Sigma=\frac{\kappa}{\eta}\rightarrow0$. At late times amount of
energy density tends to zero, so $\kappa\rightarrow0$.\\

$(ii):$ $(1+\omega)<0$,\quad
$0<{\rho_\Lambda}^{\frac{1}{n}}<\rho^{\frac{1}{n}}$,\quad
${B^{\frac{1}{n}}}>0$,\\
\begin{equation}
\kappa=\kappa_0\left({\frac{{\Omega}-{\Omega_\Lambda}}{\Omega}}\right)^{\frac{n}{|1+\omega|}}.
\end{equation}
Here, like case (i), in the early times the energy density is
infinite, i.e. $\Omega\rightarrow\infty$. So the shear scales
close to a constant value ($\kappa\rightarrow{\kappa_0}$),
therefore, the initial singularity is again isotropic and it is
dominated by the energy density, so
$\Sigma=\frac{\kappa}{\eta}\rightarrow0$. The energy density
close to an effective cosmological constant value at late times,
as a result the shear scales tends to zero, i.e.
$\Omega\rightarrow{\Omega_\Lambda}$, $\kappa\rightarrow0$.\\

$(iii):$ $(1+\omega)<0$,\quad
$0<{\rho}^{\frac{1}{n}}<{\rho_\Lambda^{\frac{1}{n}}}$,\quad
${B^{\frac{1}{n}}}<0$,\\
\begin{equation}
\kappa=\kappa_0\left({\frac{{\Omega_\Lambda}-{\Omega}}{\Omega}}\right)^{\frac{n}{|1+\omega|}}.
\end{equation}
In this case, the energy density decays in the early times and
grows with pass time. Hence, the fluid behavior is like phantom.
In the early times, energy density value tends to zero
($\Omega\rightarrow0$), so any shear component always comes to
dominate in the initial singularity, then initial singularity is
anisotropic. The energy density close to an effective
cosmological constant value at late times. Therefore, according
to Eq.(25) the shear scales tends to zero, i.e.
$\Omega\rightarrow{\Omega_\Lambda}$, $\kappa\rightarrow0$.

\section{Conclusion}

In this paper we have studied the effects of the general
non-linear term of EoS to suppressing anisotropy for models with
singularity in the early times. In this respect the our motivation
for adding a general non-linear term to EoS is that in the
context of brane world scenario the quadratic term appears in the
effective 4-dimensional equation of motion. In Section (2), we
have introduced a general form of the EoS which it include a
non-linear term. We have given the energy density as a function
of the scale factor. In Section (3), we have investigated one
specific example of the non-linear EoS, then we have summarized
the classification of the energy density evolution in three
different classes. We show that for special cases, the effect of
non-linear term of EoS isotropize the singularity at early times
and for another case this model has a phantom like behavior, then
in this case the singularity is not isotropized.


\begin{thebibliography}{99}


\bibitem{1} R. Brandenberger, arxiv: hep-th/1003.1745v1 (2010).

\bibitem{2} A. G. Riess \textit{et al}., Astron. J. {\bf116}, 1009, (1998); S. Perlmutter
\textit{et al}., Astrophys. J. {\bf517}, 565, (1999); A. G. Riess
\textit{et al}., Astrophys. J. {\bf607}, 665, (2004).

\bibitem{3} J. L. Tonry \textit{et al}., Astrophys. J. {\bf594}, 1, (2003).

\bibitem{4}  R. A. Knop \textit{et al}., Astrophys. J. {\bf598}, 102, (2003) ; B. J. Barris, \textit{et al}., Astrophys. J. {\bf602}, 571, (2004).

\bibitem{5} N. Bahcall, J. P. Ostriker, S. Perlmutter, P. J. Steinhardt,
Science {\bf284}, 1481 (1999).

\bibitem{6} C. L. Bennett \textit{et al}., Astrophys. J. Suppl. {\bf148}, 1,
(2003); L. Page \textit{et al}., Astrophys. J. Suppl. {\bf148},
233, (2003).

\bibitem{7} H. V . Peiris \textit{et al}.,  Astrophys. J. Suppl. {\bf148}, 213,
(2003).

\bibitem{8} D. N. Spergel \textit{et al}., Astrophys. J. Suppl. {\bf170}, 377, (2007).

\bibitem{9} R. Scranton \textit{et al}., arxiv: astro-ph/0307335v2,
(2003); M. Tegmark \textit{et al}., Phys. Rev. {\bf D 69}, 103501,
(2004).

\bibitem{10} I. Zlatev, L. Wang, P. J. Steinhardt, Phys. Rev. Lett. {\bf82}, 896,
(1999); P. J. Steinhardt, L. Wang, I. Zlatev, Phys. Rev. {\bf D
59}, 123504, (1999); M. S. Turner, Int. J. Mod. Phys. {\bf A
17S1}, 180, (2002); V. Sahni, Class. Quant. Grav. {\bf19}, 3435,
(2002).

\bibitem{11} R. R. Caldwell, M. Kamionkowski, N. N. Weinberg,  Phys. Rev. Lett. {\bf91}, 071301, (2003);
 R. R. Caldwell,  Phys. Lett. {\bf B 545}, 23, (2002); P. Singh, M. Sami, N. Dadhich, Phys. Rev. {\bf D 68} 023522, (2003);
  J. G. Hao, X. Z. Li,  Phys.Rev. {\bf D 67}, 107303, (2003).

\bibitem{12} A. Picon, T. Damour, V. Mukhanov, Phys. Let. B. {\bf458}, 209, (1999); M. Malquarti, E. J. Copeland, A. R. Liddle,
M. Trodden, Phys. Rev. {\bf D 67}, 123503, (2003); T. Chiba, Phys.
Rev. {\bf D 66}, 063514, (2002).

\bibitem{13} N. Breton, J. L. Cervantes-Cota, M. Salgado, The early universe
and observational cosmology, (2004).

\bibitem{14} C. B. Collins, S. W. Hawking, Astrophys. J. {\bf180}, 317, (1973).

\bibitem{15} J. D. Barrow, D. H. Sonoda, Phys. Reports {\bf139}, 1, (1986).

\bibitem{16} T. Shiromizu, K. i. Maeda, M. Sasaki, Phys. Rev. {\bf D
62}, 024012, (2000).

\bibitem{17} H. A. Bridgman, K. A. Malik, D. Wands, Phys. Rev.
{\bf D 65}, 043502, (2002).

\bibitem{18} D. Langlois, Astrophys. Space Sci. {\bf283}, 469,
(2003).

\bibitem{19} R. Maartens, Living Rev. Rel. {\bf7}, 7, (2004), [arXiv:gr-qc/0312059].

\bibitem{20} K. N. Ananda, M. Bruni, Phys. Rev. {\bf D 74}, 023524 , (2006).

\bibitem{21} R. R. Caldwell, Phys. Lett. {\bf B 545}, 23, (2002).

\bibitem{22} D. Eardley, E. Liang, R. Sachs, J. Math. Phys. {\bf13}, 99, (1972).

\bibitem{23} L. Randall, R. Sundrum, Phys. Rev. Lett. {\bf83}, 3370, (1999).

\bibitem{24} L. Randall, R. Sundrum, Phys. Rev. Lett. {\bf83}, 4690, (1999).

\bibitem{25} P. Singh, K. Vandersloot, Phys. Rev. {\bf D 72}, 084004, (2005).

\bibitem{26} A. Ashtekar, T. Pawlowski, P. Singh, Phys. Rev. {\bf D 74}, 084003, (2006).

\bibitem{27} P. Singh, K. Vandersloot, G. V. Vereshchagin, Phys. Rev. {\bf D 74}, 043510, (2006).

\bibitem{28} P. Singh, Phys. Rev. {\bf D 73}, 063508, (2006).

\bibitem{29} J. E. Lidsey, D. J. Mulryne, Phys. Rev. {\bf D 73}, 083508, (2006).

\bibitem{30} P. Singh, M. Sami, N. Dadhich, Phys. Rev. {\bf D 68}, 023522, (2003).

\bibitem{31} K. N. Ananda, M. Bruni, Phys. Rev. {\bf D 74}, 023523, (2006).


\bibitem{32} S. Nojiri, S. D. Odintsov, S. Tsujikawa, Phys. Rev. {\bf D
71}, 063004, (2005).

\end{thebibliography}
\end{document}